\documentclass[reprint,amsmath,amssymb,aps]{revtex4-2}
\usepackage{graphicx}
\usepackage{dcolumn}
\usepackage{bm}

\begin{document}
	
	\title{Entanglement recycled quantum key distribution scheme without sifting over arbitrary long distance}
	\author{Hao Shu}
	\email{hb.shu-Hao@yahoo.com}
	\affiliation{College of Mathematics, South China University of Technology, Guangzhou, 510641, P. R. China}
	\date{\today}
	
	\begin{abstract}
		Quantum key distribution(QKD) is an important area in quantum information theory. Nowadays, there are many protocols such as BB84 protocol, Lo-Chau's protocol and GR10 protocol. They usually require legitimated parties have the ability to create particles, using a sifting procedures (BB84, GR10), or must destroy entangled states (Lo-Chau). In this paper, we give a QKD scheme which can recycle entangled states and need not to run sifting procedures. The protocol use teleportation and mutual unbiased bases of qudits. Moreover, The scheme can be modified to add a third party who assumes all the states creating procedures and so the communicated parties need not to create states. This is in fact an entanglement distribution protocol. Also, the protocol can be modified for distributing key over arbitrary long distance. We compare our protocol with the previous protocols and discuss the security of it by corresponding to BB84 protocol.
	\end{abstract}
	
	\maketitle
	
	\section{Introduction}
	In the information era, cryptography is one of the most important subjects. In classical information theory, the most useful cryptosystem is RSA system of which the security is based on the low capacity of classical computers and arithmetics on factorizing large integers, but not on physical or mathematical laws. However, when quantum arithmetic are taken into amount, the system is not remain security\cite{S1994Algorithms}. On the other hand, if extending our view to quantum information theory, we may discover cryptographic protocols of which the securities only rely on physical laws and can be proven in mathematics.
	
	Suppose that two legitimated parties, says Alice and Bob, want to have a secure communication with each other. The most safe way for them is to share a one-time pad, a string of numbers they agree with but private among others. The string is called a key and only be used once. Thus, all problem here is how to transmit such a key securely. This is impossible if Alice and Bob share nothing initial. However, let us assume that Alice and Bob have an authenticated classical channel which might not be private, and a quantum channel without any further assumption. Here, we mean that if there is an eavesdropper, says Eye, then she may eavesdrop classical communications but with no abilities to forge the message or pretend to be one of the legitimated parties, while he can do anything under physical laws in the quantum channel including intercepting and resending particles. The security is in the meaning that if the eavesdropper get enough information of the key, then she is detectable by the legitimated parties and so they can abort the key. In above assumptions, we would like to discuss how legitimated parties can share a key securely by using quantum information theory. This is called quantum key distributing, and QKD for short.
	
	After quantum effects were studied, several QKD protocols were published. The first protocol may be the famous BB84 protocol, which was published by Bennett and Brassard in 1984\cite{BB1984Quantum}. The protocol in fact use a sifting method on two mutual unbiased basis in $C^2$. This procedure waste half of states in average to obtain security. There is an equivalent version called BBM92 protocol using EPR particles\cite{BB1992Quantum}. Also, BB84 protocol can be extended to use more mutual unbiased basis and in $C^d$ such as so called six-states protocol\cite{B1998Optimal}, and \cite{CB2002Security}. Such protocol might increasing security but decreasing efficient since sifting procedures. Other protocols including Ekert's protocol based on Bell's theorem\cite{E1991Quantum}, Lo-Chau's protocol based on distributing EPR pairs\cite{LC1999Unconditional}, Shor-Preskill's protocol based on CSS codes\cite{SP2000Simple} and GR10 protocol based on probatilistic teleportation\cite{GR2010Quantum}.
	
	The BB84 protocol might be the most researched protocol. A secure proof can be found in\cite{SP2000Simple}. The main flaw of BB84 protocol may be the necessity of sifting procedures which waste half of the states\cite{BB1984Quantum}. The same flaw also happen in extended versions of BB84 protocol\cite{BB1992Quantum}\cite{B1998Optimal}\cite{CB2002Security} and in GR10 protocol\cite{GR2010Quantum}. Shor-Prestill's CSS protocol\cite{SP2000Simple} and Lo-Chau's entangled based protocol\cite{LC1999Unconditional} need not to tolerate such a procedure. However, the CSS protocol\cite{SP2000Simple} require one of the legitimated parties has the ability to create states, this requirement also necessary in Ekert's protocol\cite{E1991Quantum}, and the entangled-based protocol\cite{LC1999Unconditional}\cite{SP2000Simple}, besides, need to destroy entangled states. An other problem is practical, the distance of transmitting a states might be limited since the attenution of single and this can restrict the distance of key distribution. Thus, a discussion of distribution distance is significant.
	
	In this paper, we give a protocol based on teleportation and mutual unbiased basis in $C^d$, which need not to tolerate sifting procedures and can recycle entangled states. That means, in theoretical, only one maximally entangled state is needed. Moreover, after a modify, the protocol can be extended by add a third party, says Charlie, who assumes all the creating of the states. This is in fact an entanglement distribution protocol with third party and, similarly, all entangled states can be recycled. We also mention that like in entangled-based protocols, an entanglement distill procedure might be supplied for getting nearly perfect maximally entangled states. Moreover, the protocol can be modified for distributing a secret key over arbitrary long distance.
	
	\section{A review of teleprotation and mutual unbiased basis}
	In this section, let us have a review of teleprotation and mutual unbiased basis.
	
	Teleprotation protocol was firstly published in 1993, using EPR pairs\cite{BB1993Teleporting}. Let us give a description in $C^d\otimes C^d$. Suppose that Alice and Bob share a maximally entangled state, says $|\varphi \rangle=\frac{1}{\sqrt{d}}\sum_{j=0}^{d-1}|j\rangle_{A}|j\rangle_{B}$, where $\left \{ |j\rangle |j=0,1,...,d-1 \right \}$ is an orthonormal base in $C^d$. Alice wants to teleport a state, says $|\phi\rangle=\sum_{s=0}^{d-1}a_{s}|s\rangle$, where $\sum_{s=0}^{d-1}|a_{s}|^2=1$, to Bob. The procedure can be described as follow. Alice consider the states together, which can be written as $|\phi\rangle_{A'}|\varphi \rangle_{AB}=\frac{1}{\sqrt{d}}\sum_{s=0}^{d-1}a_{s}|s\rangle_{A'}\sum_{j=0}^{d-1}|j\rangle_{A}|j\rangle_{B}=\frac{1}{\sqrt{d}}\sum_{k,l=0}^{d-1}|\psi_{k,l}\rangle_{A'A}\sum_{s=0}^{d-1}a_{s}e^{-k(s+l)}|s+l\rangle_{B}$, where $e$ is a primitive d-th root of unity and $|\psi_{k,l}\rangle=\frac{1}{\sqrt{d}}\sum_{j=0}^{d-1}e^{(i+l)k}|j\rangle_{A}|j+l\rangle_{B}=(I\otimes Z^{k}X^{l})\psi_{0,0}$ be generalized Bell states and X, Z be generalized Pauli operators. The set $\left \{|\psi_{k,l}\rangle|k,l=0,1,...,d-1\right \}$ form an orthonormal base in $C^d\otimes C^d$. Now, Alice measures via this base on her partite $A'A$ and gets an outcome, says (k,l). Hence, Bob has a state equal $\sum_{s=0}^{d-1}a_{s}e^{-k(s+l)}|s+l\rangle$. Then Alice can announce (k,l) and Bob can do a local transformation to get $|\phi\rangle$.
	
	Two orthonormal basis $\beta_{1}$, $\beta_{2}$ in $C^d$ are said to be mutual unbiased if for any $|\gamma_{1}\rangle\in\beta_{1}$, $|\gamma_{2}\rangle\in\beta_{2}$, we have $|\langle \gamma_{1}|\gamma_{2}\rangle|=\frac{1}{d}$. A set of orthonormal basis are said to be mutual unbiased if any two of them are all mutual unbiased. Of course, for every non-trivial system, there are two mutual unbiased basis. For example, let $\beta_{1}=\left \{ |j\rangle |j=0,1,...,d-1 \right \}$, $\beta_{2}=\left \{ |\alpha_{j}\rangle |j=0,1,...,d-1 \right \}$, where $|\alpha_{j}\rangle=\frac{1}{\sqrt{d}}\sum_{i=0}^{d-1}e^{ij}|i\rangle$. We know that if d is a prime power, then there are $d+1$ mutual unbiased basis in $C^d$\cite{BB2008A}.
	
	\section{The QKD protocol}
	We will give a QKD scheme in this section. The scheme is based on teleportation and mutual unbiased basis in $C^d$. The protocol has ten steps as follow:
	
	Step 1: Alice and Bob agree to encode $0,1,...,d-1$ by states $|0\rangle, |1\rangle,..., |d-1\rangle$ in $C^d$, where of course $d\geq 2$, and they agree with unitary operators $U_{i}$, where $i=0,1,..., m-1$, such that $U_{i}S=\left \{ U_{i}|j\rangle |j=0,1,...,d-1 \right \}$ are mutual unbiased basis in $C^d$. Hence, $S=\left \{ |j\rangle |j=0,1,...,d-1 \right \}$ and let $U_{0}=I$. Of course, let $m\geq 2$ which is suitable since there are at least two mutual unbiased basis in a non-trivial system.
	
	Step 2: To share a N-bit key string, Alice create 2N maximally entangled states, all be $|\psi\rangle=\frac{1}{\sqrt{d}}\sum_{j=0}^{d-1}|j\rangle_{A}|j\rangle_{B}$.
	
	Step 3: Alice choose a string $b=b_{1}b_{2},...,b_{2N}$ with $b_{r}\in \left \{ 0,1,...,m-1 \right \}$ randomly and transform the r-th state by $I\otimes U_{b_{r}}$ such that the state become $|\psi_{r}\rangle=\frac{1}{\sqrt{d}}\sum_{j=0}^{d-1}|j\rangle_{A}U_{b_{r}}|j\rangle_{B}$
	
	Step 4: Alice send partite B of the 2N states to Bob.
	
	Step 5: After receiving the particles, Bob publicly announce this fact.
	
	Note that Alice and Bob can do these steps before they want to share a secret key, when they think that the quantum channel might not be very noisy.
	
	Step 6: Alice chooses a randomly 2N string of 0,1,..., d-1, creates corresponding states in S (like in BB84 protocol and can be replaced by measuring maximally entangled states as in BBM92 protocol). Alice teleport these 2N states by the 2N maximally entangled states. If Alice teleport state $|s_{r}\rangle$ and gets an outcome, says $(k_{r},l_{r})$, after measuring via basis $\left \{|\psi_{k,l}\rangle|k,l=0,1,...,d-1\right \}$ in the r-th position, then Bob's state should be $e^{-k_{r}(s_{r}+l_{r})}U_{b_{r}}|s_{r}+l_{r}\rangle$.
	
	Step 7: Alice publicly publish the string $l=l_{1},l_{2},...,l_{2N}$, and the string $b=b_{1},b_{2},...,b_{2N}$.
	
	Step 8: Bob transform his r-th state by $U_{b_{r}}^{-1}$ since now he knows the string $b$.
	
	Step 9: Bob measures his states via base S. His result should be $s_{r}+l_{r}$ in the r-th position if there is no noisy. Since he knows the string $l$, he can get the value of $s_{r}$.
	
	Step 10: Alice or Bob chooses a random N-string of the 2N string as check bits. They publish their values on these bits and estimate the error rate. If the error rate is acceptable, they declare that there is no eavesdroppers and use the remain N string as a (raw) secret key. If not, the abort the string. This procedure and the rest of the procedures such as error correcting and private amplification are analog to BB84 protocol.
	
	\section{Analyse of the protocol}
	Before given a secure proof, let us give an analyse of the protocol.
	
	The main advantage of our protocol, compare with BB84-like protocols\cite{BB1984Quantum}\cite{BB1992Quantum}\cite{B1998Optimal}\cite{CB2002Security}\cite{SP2000Simple} or GR10-like protocols\cite{GR2010Quantum}\cite{LR2020Asymptotic}, is that our protocol need not suffer a sifting procedure and so the efficiency will be twice as in such protocols and can use more mutual unbiased basis to increase security without decreasing efficiency. Note that the argument can be extended to probatilistic teleprotation case like GR10 protocol\cite{GR2010Quantum} but without sifting entangled states.
	
	The method of avoiding sifting procedures has been used in Lo-Chau's and Shor-Preskill's protocol with d=m=2\cite{LC1999Unconditional}\cite{SP2000Simple}. However, while Lo-Chau's protocol\cite{LC1999Unconditional} has to destroy maximally entangled states, our protocol can recycle maximally entangled states. We mention that after the procedures of our protocol, Alice gets one of the generalized Bell states and can do a transformation making it be the standard one. She can reuse the state and thus, in theoretical, Alice only need to create one maximally entangled state.
	
	One the other hand, Alice and Bob may share maximally entangled states (Step 1 to Step 5) before they want to create a secret key when they think the quantum channel is not noisy enough.
	
	Of course, like a entangled based protocol, Alice and Bob may prefer to check maximally entangled states instead of checking the final bits. We mention that a checking procedure can run before step 6 to check whether they exactly share two partite maximally entangled states and not purified by an eavesdropper. Just let Alice measure a random N bit of the 2N string, each via base $U_{i}S$ by choosing N $U_{i}$ randomly and send the choice of check bits, the measuring result and string b to Bob. Then Bob corrects his state by string b and measuring the check bits using the same basis as Alice did. They compare their results to see if the error rate is acceptable. This can be a substitution of checking procedure in step 10. However, this procedure need to consume entanglement and Alice can recycle N maximally entangled states but not 2N.
	
	Another advantage of our protocol is that the success rate of the protocol is mainly depend on the situation of maximally entangled states that Alice and Bob share. If they share perfect bipartite maximally entangled states, the protocol can work well. If not, they may choose to run a entanglement distill protocol to get nearly perfect maximally entanglement states then run the protocol (Step 6 to Step 10).
	
	As we have mentioned, the protocol can be modified to add a third party who assumes all the creating of states such that Alice and Bob need not to have the ability of creating states. This is another reason why the protocol is considered to be valuable. We will discuss this after giving a secure proof.
	
	Also, the protocol an be modified for distributing a secret key over arbitrary long distance. We will discuss this by adding parties between Alice and Bob.
	
	However, there is a main disadvantage. Like in Lo-Chau's and Shor's protocol, the protocol require quantum memories. Alice and Bob have to keep their states until Bob get the string b. But the discussion of hardware is beyond this paper.
	
	\section{Security}
	Let us consider that there is an eavesdropper, says Eye, who want to steal the secret key of Alice and Bob. The only chance she can do is affecting the entangled distribution procedure and getting classical messages  Alice and Bob communicating in public channel, since other procedures and all local. She has two kinds of attacks. She can intercept the B partite of maximally entangled states, then either takes it herself and send to Bob another state instead (Let us call this kind of attack a substituted attack), or add an auxiliary partite (her partite) then do a transformation, resend the B partite to Bob (Let us call it a purified attack).
	
	Let us analyse the substituted attack firstly. We an show that if Eye implement such an attack, then Bob gets right result (the result as he should agree with Alice) with probability $\frac{1}{d}$, totally randomly. That means of course Eye is detectable when Alice and Bob run the checking procedure. To see this, only notice that Bob and Eye has no communication at all after Bob gets the state and in this time which states Alice will teleport is completely random. If Eye can control the state of Bob even if after she get the state teleport by Alice, then Eye and Bob can communicate faster then light speed. This is a contradiction with relativity theory.
	
	We can verify this for our assumption. If Eye implement such an attack, says Eye steal the B partite of $|\psi_{r}\rangle$ and send Bob the B partite of a state, says $|\eta\rangle_{BC}$, cerated by her. Now after Alice teleporting a state, says $|s\rangle$, and publishing string b and string $l$, eye exactly get the right state. However, the only thing Bob and Eye share is $|\eta\rangle_{BC}$, and the conclusion is followed by the proposition.
	
\textbf{Proposition}: Let Bob and Eye share a state $|\eta\rangle_{BC}$ and Eye choose a state $|s\rangle$ in $S=\left \{ |j\rangle |j=0,1,...,d-1 \right \}$ randomly. They are freely to use local resource but have no channels. Let Bob get a state. Then Bob's state is the same as Eye's is of probability $\frac{1}{d}$ in average.

\textbf{Proof}: All Bob and Eye can do are local transformations and local measurements. Since a local unitary operator composite a local measurement is in fact a local measurement, without loss generality, assume that Bob provide a local POVM $\left \{ \Pi_{t}\right \}_{t=0,1,...,d-1}$, and guess Eye's state is $|t\rangle$ if his outcome is t, while Eye provide a POVM $\left \{ M_{k}\right \}_{k=0,1,...}$. The final POVM then become $\left \{ \Pi_{t}\otimes M_{k} \right \}_{t=0,1,...,d-1 and k=0,1,...}$. The probability of resulting outcome to be (t,k) is $P_{t,k}=\langle\eta|\Pi_{t}\otimes M_{k}|\eta\rangle$. The probability of Bob getting t is $P_{t}=\sum_{k}\langle\eta|\Pi_{t}\otimes M_{k}|\eta\rangle$ and the probability of Bob's outcome is right is $P_{r}=\sum_{t=0}^{d-1}\frac{1}{d}\sum_{k}\langle\eta|\Pi_{t}\otimes M_{k}|\eta\rangle=\frac{1}{d}$.	
$\hfill\blacksquare$
	
	Now, let us consider the other kind of attack, says the purified attack. The security of the protocol is the same as BB84 protocol. To see this, let us make a correspondence of the protocol and BB84 protocol\cite{BB1984Quantum}\cite{SP2000Simple}. Assume that Eye add an auxiliary partite, says E, then do a transformation, says $U_{E}$ and resend the B partite to Bob. Since Eye only have entrance of partite B and E (partite A is never sent), $U_{E}$ is an unitary operator on partite B and E. Note that after such an attack for BB84 protocol, Bob's state become $U_{E}\left [ U_{i}|s\rangle_{B}|0\rangle_{E}\right ]$, if Alice send the state $U_{i}|s\rangle$, where $i=0,1$, $U_{0}=I,
	U_{1}=
	\begin{bmatrix}
		1 & 1
	 \\ 1 & -1
	\end{bmatrix}$
	via the computational base.
	As for our protocol, after Alice's teleportation, an easy calculation shows that, Bob's state will be $e^{-k(s+l)}U_{E}\left [ U_{i}|s+l\rangle_{B}|0\rangle_{E}\right ]$ if Alice send $|s\rangle$ by the maximally entangled state whose  B partite is transformed by the operator $U_{i}$. The global phase is not important and since $l$ is published publicly, Bob's state in both of the protocols are exactly equivalent. This shows that the protocol is as secure as BB84 protocol\cite{BB1984Quantum}\cite{SP2000Simple}.
	
	\section{Quantum key distribution with a third party}
	Suppose now Alice and Bob are unable to create states, but a third party, says Charlie, can. Can Alice and Bob share a secret key with the help of Charlie without letting Charlie know the key? The answer is yes. We can modify our protocol to accomplish this. We only discuss for qubit and three parties case. It can be easily generalized to general cases such as qudit, more parties and probatilistic teleportation cases.
	
	Let Charlie create 2N state, all be $|GHZ\rangle=\frac{1}{\sqrt{2}}\sum_{j=0}^{1}|j\rangle_{C}|j\rangle_{A}|j\rangle_{B}$ and sending partite A, B to Alice and Bob, respectively. After Alice and Bob declare that they have received their particle. Charlie begin a  protocol, teleporting 2N states, all be $|+\rangle_{C_{1}}|+\rangle_{C_{2}}$, where $|+\rangle=\frac{1}{\sqrt{2}}\sum_{j=0}^{1}|j\rangle$ using those $|GHZ\rangle$. We calculate as follow. Let
	
    $|a\rangle=\frac{1}{\sqrt{2}}(|0\rangle|0\rangle|0\rangle+|1\rangle|1\rangle|1\rangle)$,
    \\	
	$|b\rangle=\frac{1}{\sqrt{2}}(|0\rangle|0\rangle|1\rangle+|1\rangle|1\rangle|0\rangle)=(I\otimes I\otimes X)|a\rangle$,
	\\
	$|c\rangle=\frac{1}{\sqrt{2}}(|0\rangle|0\rangle|0\rangle-|1\rangle|1\rangle|1\rangle)=(I\otimes Z\otimes I)|a\rangle$,
	\\
	$|d\rangle=\frac{1}{\sqrt{2}}(|0\rangle|0\rangle|1\rangle-|1\rangle|1\rangle|0\rangle)=(I\otimes Z\otimes X)|a\rangle$,
	\\
	$|e\rangle=\frac{1}{\sqrt{2}}(|0\rangle|1\rangle|0\rangle+|1\rangle|0\rangle|1\rangle)=(I\otimes X\otimes I)|a\rangle$,
	\\
	$|f\rangle=\frac{1}{\sqrt{2}}(|0\rangle|1\rangle|1\rangle+|1\rangle|0\rangle|0\rangle)=(I\otimes X\otimes X)|a\rangle$,
	\\
	$|g\rangle=\frac{1}{\sqrt{2}}(|1\rangle|0\rangle|0\rangle-|0\rangle|1\rangle|1\rangle)=(I\otimes ZX\otimes X)|a\rangle$,
	\\
	$|h\rangle=\frac{1}{\sqrt{2}}(|1\rangle|0\rangle|1\rangle-|0\rangle|1\rangle|0\rangle)=(I\otimes Z\otimes I)|a\rangle$.
	
	where X, Z are pauli operators, and these states form an orthonormal base in $C^2\otimes C^2\otimes C^2$. Now,
	
\begin{widetext}
\begin{equation}
\begin{aligned}
		& |+\rangle_{C{1}}|+\rangle_{C{2}}|GHZ\rangle_{CAB} \\
		=& \quad\frac{1}{4}\left[ (|a\rangle+|c\rangle)_{C_{1}C_{2}C}|0\rangle_{A}|0\rangle_{B}+(|b\rangle+|d\rangle)_{C_{1}C_{2}C}|1\rangle_{A}|1\rangle_{B}+(|e\rangle-|h\rangle)_{C_{1}C_{2}C}|0\rangle_{A}|0\rangle_{B}+(|f\rangle-|g\rangle)_{C_{1}C_{2}C}|1\rangle_{A}|1\rangle_{B}\right]
       \\
       & +  \frac{1}{4}\left[(|f\rangle+|g\rangle)_{C_{1}C_{2}C}|0\rangle_{A}|0\rangle_{B}+(|e\rangle+|h\rangle)_{C_{1}C_{2}C}|1\rangle_{A}|1\rangle_{B}+(|b\rangle-|d\rangle)_{C_{1}C_{2}C}|0\rangle_{A}|0\rangle_{B}+(|a\rangle-|c\rangle)_{C_{1}C_{2}C}|1\rangle_{A}|1\rangle_{B}\right]
		\\
		=& \frac{1}{4}\left [ (|a\rangle+|b\rangle+|e\rangle+|f\rangle)_{C_{1}C_{2}C}(|0\rangle|0\rangle+|1\rangle|1\rangle)_{AB}+(|c\rangle-|d\rangle+|g\rangle-|h\rangle)_{C_{1}C_{2}C}(|0\rangle|0\rangle-|1\rangle|1\rangle)_{AB}\right].
\end{aligned}
\end{equation}
\end{widetext}
	
	Let Charlie measures via the above base, called it $S'$, and get an outcome. He will get 2N states in $S'$ and he can transform these states to the standard $|GHZ\rangle$ by Pauli operators, recycling the states. On the other hand, Alice and Bob now share  maximally entangled states, say $\frac{1}{\sqrt{2}} (|0\rangle|0\rangle+|1\rangle|1\rangle)$ or $\frac{1}{\sqrt{2}}(|0\rangle|0\rangle-|1\rangle|1\rangle)$. Now Alice and Bob should check whether they actually share such maximally entangled states. The method have been stated in the Analyse section. Alice randomly choose N states as check bits, choose a N string of 0 and 1, denoted by $b=b_{1},...,b_{N}$, randomly and measure the r-th check bits via base $\left \{ |0\rangle, |1\rangle \right \}$ if $b_{r}=0$, and via base $\left \{ |+\rangle, |-\rangle \right \}$ if $b_{r}=1$. She announce which states did she consider as check bits and the string b, together with her measuring result. Then Bob measures his partite of the check bits via the same base and estimate whether the error rate is acceptable. The remaining procedure is the same as in the Analyse section.
	
	If Charlie is trusted, they can also use the three parties protocol as similarly as the ordinary one. Says, Charlie provide a Hamard gate or I(identity operator) on partite A and B of the $|GHZ\rangle$ states before sending the partite to Alice and Bob, randomly. After they receiving the particles, he published the operators he chose for partite A and B, publicly. Alice and Bob then provide an inverse transformation to get their states. The rest of the procedures is the same as in the ordinary protocol. Alice and Bob can check the entangled state firstly, or they can measure their partite to get a secret key as in Lo-Chau's protocol\cite{LC1999Unconditional} or use the states to teleportation as in our protocol, then check the result.
	
	We mention that the above protocol is in fact an entanglement distribution protocol. Thus, supplements of entanglement distribution protocols may work. For example, a entanglement distill procedure maybe implemented if needs.
	
	Of course, in the above protocol, Charlie may choose to teleport only $|+\rangle$ instead of $|+\rangle|+\rangle$. However, if so, Charlie can only recycle a bipartite maximally entangled states but not a $|GHZ\rangle$ state. Hence, we use a product state to save a entangled state.
	
	\section{Quantum key distribution over arbitrary long distance}
	In practise, signal attenuation might be a significant problem. It limits transmitted distance of states. In our protocol, the distance of QKD is limited by transmitted distance of B partite of maximally entangled states. However, the protocol can be modified for arbitrary long distance.
	
	Let us give a equivalent statement of our protocol. Alice and Bob share maximally entangled states, says $|\psi\rangle$. This procedure is similar as in the protocol but without transforming via $U_{i}$. Instead of transforming maximally entangled states, Alice now transform states in S. That is when Alice wants to teleport $|s\rangle$, she teleport $U_{i}|s\rangle$ instead by a random $U_{i}$ as in the protocol. After teleporting states to Bob, Alice publish all $U_{i}$ she chose, that is the string $b$. She also publish the string $l$. Then Bob correct his states via string $b$, measuring via base S and correct his results via string $l$ as in the protocol. Others procedures such as checking is the same as in the protocol.
	
	In order to distribute key string over arbitrary long distance, we should add several parties between Alice and Bob. Let us assume that a maximally entangled state can be distributed over a distance, says D, without a significant loss. To distribute a key string over a distance ND, let us add $N-1$ parties, say $E_{1},...,E_{N-1}$, between Alice and Bob such that every alternated parties have a distance D. That is , let $E_{i}$ and $E_{i+1}$ have a distance D for $i=0,1,...N-1$ and here, let $E_{0}$ be Alice and $E_{N}$ be Bob. Now maximally entangled states can be distributed over $E_{i}$ and $E_{i+1}$. This procedure can be done by requiring average half of the parties have ability to create maximally entangled states. The remain procedure is to teleport transformed states party by party (in fact, only parties with even label need to  teleport states and the odd-label parties only need to prepare maximally entangled states and send each partite of them to alternated two parties, respectively), from $E_{0}$(Alice) to $E_{N}$(Bob), each party publishes his measuring outcome (only need the $l$ string of each parties) . After Bob receiving the states, Alice publish the string $b$ and Bob correct his states by $b$, measuring via base S and correct results via string $l$ of all other parties. The remain procedures are similar to the protocol. Note that by adding enough parties, the distribution distance can be arbitrary long (N can be large), in theoretical.

	\section{Conclusion}
	In this paper, we give a quantum key distribution scheme based on teleportation and mutual unbiased basis of $C^d$. Our protocol, on one hand, need not to suffer a sifting procedure and so save half of the states, and on the other hand, recycle maximally entangled states. We give an analyse of our protocol, compare it with several protocols and give a secure proof. Moreover, our protocol can be modified by add a third party who assumes all state creating. We also demonstrated that our protocol can be modified for distributing a secret key over arbitrary long distance. These give other values of our protocol.
	
    As for remaining problems. One might generalize the protocol to probabilistic telrportation cases, to quantum secret sharing or others. Another interesting problem may be consider such a protocol with error correcting code and give a completed analyse of its efficiency. Also, one might discuss whether a modified protocol can work even if some of nodes are untrusted, not only detect them. Besides, discussions of such a scheme over a network seems to be significant. The final problem is about hardware. As we have seen in the paper, implement the protocol require quantum memories which can be a problem in practice.
	
    \bibliography{Bibliog}
	\end{document}